# Solid-phase silicon homoepitaxy via shear-induced amorphization and recrystallization

Thomas Reichenbach,[1,2] Gianpietro Moras,[1,*] Lars Pastewka,[1,3,4,5] Michael Moseler[1,2,4]

[1]*Fraunhofer IWM, MicroTribology Center µTC, Wöhlerstr. 11, 79108 Freiburg, Germany*

[2]*Institute of Physics, University of Freiburg, Hermann-Herder-Str. 3, 79104 Freiburg, Germany*

[3]*Department of Microsystems Engineering, University of Freiburg, Georges-Köhler-Allee 103, 79110 Freiburg, Germany*

[4]*Freiburg Materials Research Center, University of Freiburg, Stefan-Meier-Str. 21, 79104 Freiburg, Germany*

[5]*Cluster of Excellence livMatS, Freiburg Center for Interactive Materials and Bioinspired Technologies, University of Freiburg, Georges-Köhler-Allee 105, 79110 Freiburg, Germany*

*Corresponding author: gianpietro.moras@iwm.fraunhofer.de

**Abstract**

The development of epitaxy techniques for localized growth of crystalline silicon nanofilms and nanostructures has been crucial to recent advances in electronics and photonics. A precise definition of the crystal growth location, however, requires elaborate pre-epitaxy processes for substrate patterning. Our molecular dynamics simulations reveal that homoepitaxial silicon nanofilms can be directly deposited by a crystalline silicon tip rubbing against the substrate, thus enabling geometrically controlled crystal growth with no need for substrate pre-patterning. We name this solid-phase epitaxial growth triboepitaxy as it solely relies on shear-induced amorphization and recrystallization that occur even at low temperature at the sliding interface between two silicon crystals. The interplay between the two concomitant, shear-induced processes is responsible for the formation of an amorphous sliding interface with constant nanometric thickness. If the two elastically anisotropic crystals slide along different crystallographic orientations, the amorphous layer can move unidirectionally perpendicular to the sliding plane, causing the crystal with lowest elastic energy per atom to grow at the expenses of the other crystal. As triboepitaxial growth is governed by the shear elastic response of the two crystals along the sliding direction, it can be implemented as a mechanical scanning-probe lithography method in which epitaxial growth is controlled by tuning the crystallographic misorientation between tip and substrate, the tip's size or the normal force. These results suggest a radically new way to conceive nanofabrication techniques that are based on tribologically induced materials transformations.

The development of nanometer-accurate techniques for localized epitaxial growth of single-crystalline films and nanostructures is essential for modern electronic, optoelectronic and photonic applications.[1,2] To precisely control the location, shape and size of the grown crystalline structures, complex processes, based for instance on nanolithography,[3] are necessary to predefine the growth patterns on the substrate surface.[1] A direct deposition of crystalline patterns by means of mechanical scanning-probe lithography techniques[4] would help circumvent this issue as the deposition process would be localized at the contact between writing tip and substrate and, in addition, its accuracy would not be diffraction-limited. However, the implementation of mechanical or tribological scanning-probe nanolithography in silicon generally relies on nanoscratching to remove surface atomic layers,[5,6] or to induce localized surface amorphization[7] or oxidation.[8] Despite producing nanoscale features on silicon substrates, these techniques have not been used for localized deposition of crystalline silicon films. Yet our previous studies on shear-induced amorphization of diamond[9] and silicon[10] suggest that the direct deposition of amorphous silicon (a-Si) films from a silicon tip onto a silicon substrate might be possible. Such a-Si films could be recrystallized in a following step by thermally induced solid-phase epitaxy.[11]

In order to combine these two steps into a single mechanical process, we propose that the same plastic shear deformation that is responsible for amorphization can also enable recrystallization, since, similarly to temperature, it provides the atomic mobility necessary to overcome the energy barriers for recrystallization of the metastable amorphous phase. Indeed, as we reported recently,[10] a nanoscale amorphous layer forms at the sliding interface between two silicon crystals. A competition between shear-induced amorphization[9] and recrystallization[10] at the amorphous-crystal interfaces delimiting the a-Si region results in a constant thickness of the amorphous layer, whose position changes stochastically with no unidirectional drift if the two crystals have identical orientation. In this article, we use molecular dynamics (MD) simulations to explore conditions that render the movement of the amorphous layer unidirectional, thus leading to the growth of one crystal at the expense of the other one.

We find that this tribologically induced homoepitaxial process is possible if the two sliding crystals have different crystallographic orientations and we name it "triboepitaxy", a combination of nanotribology and solid-phase epitaxy. We show that, for a variety of surface orientations, the growth direction is determined by the elastic energy per atom $E_{el}$: the crystal with lowest $E_{el}$ grows. Hence, triboepitaxial growth can be controlled by exploiting elastic anisotropy, *i.e.* by tuning the relative orientation of the two crystals with respect to the sliding direction. Finally, we propose that triboepitaxy could enable a novel scanning-probe nanolithography technique for the deposition of crystalline silicon nanostructures with no need for elevated temperatures or contaminating species to catalyse recrystallization.[3] In this case, triboepitaxial growth can also be controlled by exploiting elastic finite-size effects, *e.g.* by tuning shape and size of the writing tip.

We perform classical reactive MD simulations[12,13] to study the evolution of a shear-induced a-Si phase between two diamond-cubic silicon crystals.[9,10] Different interatomic potentials,[13–15] including a recent potential based on machine learning of density-functional-theory results,[15,16] yield comparable results (Supporting Information S1). Figure 1a displays the interface between two silicon (110) surfaces before sliding along the [1$\bar{1}$0] direction (periodic boundary conditions are applied in the sliding plane; analogous simulations for (001) surfaces are described in Supporting Information S6.1). After imposing an external normal pressure $P_n$ and thermalizing the system to room temperature by means of specially tailored barostat and thermostat,[9] a top layer of atoms is rigidly moved at a constant velocity $v$ along the sliding direction, while a bottom layer is kept fixed.

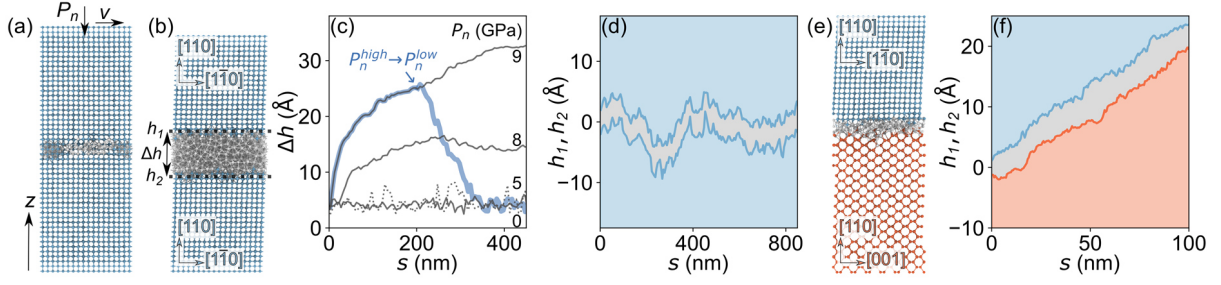

**Figure 1.** Two diamond-cubic Si(110) crystals in relative motion. (a) Interface under normal load $P_n$ before sliding along the $[1\bar{1}0]$ direction with $v = 10$ ms$^{-1}$. (b) After shearing at $P_n = 8$ GPa for a sliding distance $s = 300$ nm. Blue and gray spheres represent Si atoms in the crystalline and amorphous regions, respectively. (c) a-Si thickness $\Delta h(s, P_n)$ as a function of $s$ for different $P_n$ (dotted line: $P_n = 0$). The blue line shows $\Delta h(s)$ when the initial $P_n^{high} = 9$ GPa is reduced to $P_n^{low} = 5$ GPa at $s = 200$ nm. (d) Vertical positions $h_1(s)$ and $h_2(s)$ of the two amorphous-crystal interfaces for $P_n = 5$ GPa. Amorphous and crystalline regions are grey and blue, respectively. (e) Shearing of differently oriented crystals: the upper crystal slides along the $[1\bar{1}0]$ and the lower crystal (in red) along the $[001]$ direction. (f) $h_1(s)$ and $h_2(s)$ for the system in (e) with $P_n = 0$.

Upon sliding, an a-Si layer forms (snapshot in Figure 1b) at the interface between the two crystals. It entirely accommodates the plastic shear deformation of the system[10] and its thickness $\Delta h(s, P_n)$ increases with sliding distance $s$ and with applied normal pressure $P_n$ (Figure 1c for $P_n = 0 - 9$ GPa). As previously reported,[10] the crystalline-to-amorphous transition is mechanically induced, local temperatures remain below 400 K and a melting transition can be excluded. Interestingly, as the sliding distance increases, $\Delta h(s, P_n)$ saturates at a constant value $\Delta h_{eq}(P_n)$. Since mechanical amorphization alone would lead to a strict monotonous increase of $\Delta h$,[9,10] saturation indicates the presence of a competing process, namely shear-induced recrystallization.[10] Indeed, the vertical positions $(h_1, h_2)$ of the lower and upper amorphous-crystal interfaces fluctuate with constant $h_2 - h_1 = \Delta h$ and no clear drift (Figure 1d).

A further evidence of the recrystallization process is provided by the blue line in Figure 1c. Here, we perform a sliding simulation at high pressure ($P_n^{high} = 9$ GPa) and suddenly decrease the normal pressure to $P_n^{low} = 5$ GPa at a sliding distance $s_{P_n^{high} \to P_n^{low}} = 200$ nm. This causes a rapid decrease of the a-Si thickness from $\Delta h\left(s_{P_n^{high} \to P_n^{low}}, P_n^{high}\right) = 2.5$ nm to $\Delta h_{eq}(P_n^{low}) = 0.4$ nm, *i.e.* a 2.1-nm growth of the crystals. In accordance with the dependency $\Delta h_{eq}(P_n)$, $\Delta h_{eq}(P_n^{low})$ is exactly the a-Si thickness the system would have reached had the simulation been performed at $P_n^{low}$ from the beginning.

While no net crystal growth occurs for symmetric tribopartners (*i.e.* same crystal orientation and sliding direction), we now show that shear-induced growth of one crystal at the expense of the other can be achieved by breaking the symmetry of the sliding system. Figure 1e displays a Si(110) sliding system in which the lower crystal is rotated by 90° with respect to the upper one, so that sliding proceeds along different crystallographic directions for the two crystals. Now, the lower crystal grows rapidly, as shown by the upwards migration of $h_1$ and $h_2$ in Figure 1f, while $\Delta h$ remains roughly constant at $\Delta h_{eq}$ (Movie S1). Growth is conveniently characterized by the rate $\xi(s) = \frac{dh_1(s)}{ds}$, which is ~0.025 and approximately constant in Figure 1f.

To elucidate the mechanism underlying the observed triboepitaxial growth, we first focus on the shear stress $\tau(s)$. Although the a-Si region migrates, the time-averaged shear stress $\langle \tau \rangle$ remains constant. Since also $\xi(s)$ and $\Delta h(s)$ remain roughly constant, we conclude that the properties of the crystal-amorphous-crystal transition region remain unaltered throughout the simulation. Figure 2a shows a saw-tooth-shaped $\tau(s)$ characteristic of stick-slip instabilities. During stick phases, the system deforms elastically. When $\tau(s)$ exceeds the interfacial shear strength, the crystal-amorphous-crystal transition region undergoes plastic slip and its atoms gain mobility. In Figure 2a, we follow the typical trajectory of an atom directly involved in triboepitaxy during several stick-slip phases (Movie S2). Initially, the atom is part of the upper crystal. A series of plastic slip events cause the atom to be dragged into the a-Si region and to move to the lower amorphous-crystal interface where it eventually recrystallizes.

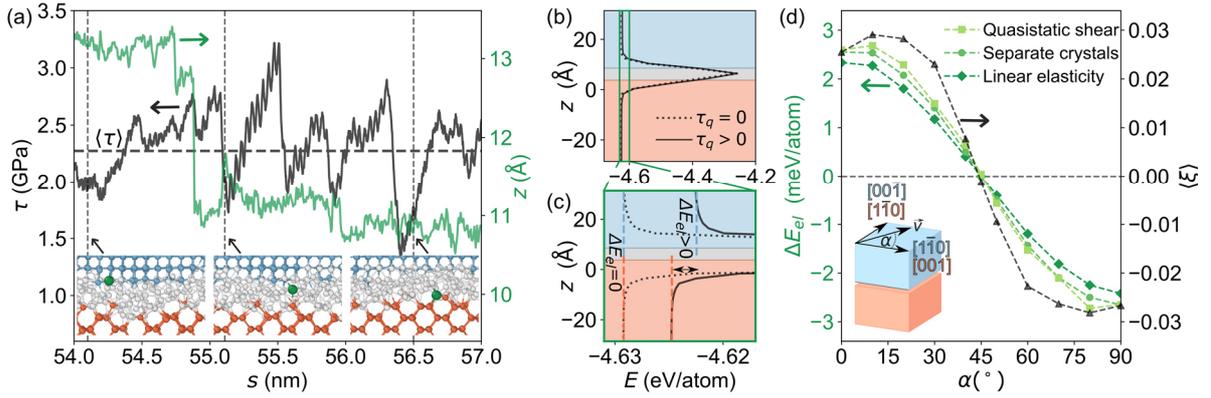

**Figure 2.** Triboepitaxial growth mechanism. (a) Running-average of shear stress $\tau$ (black) during the MD simulation of Figure 1f, showing stick-slip instabilities. The horizontal dashed line represents $\langle\tau\rangle$. The green line is the $z$-coordinate of an atom (green sphere) that is transferred from the upper to the lower crystal as a result of triboepitaxy. (b) Potential energy per atom $E$ as a function of $z$ for applied shear stress $\tau_q = 0$ and $\tau_q > 0$. (c) Zoom-in with vertical dashed lines highlighting the energy levels of the crystals. (d) Elastic energy differences per atom $\Delta E_{el}$ (green) between upper and lower crystal when the original sliding direction (Figure 1e) is rotated by an angle α (inset). $\Delta E_{el}$ is determined by quasistatically shearing either the tribosystem (squares) or two single crystals (circles) and by linear elasticity (diamonds). Black triangles show $\langle\xi\rangle$ measured in MD simulations.

To understand what drives the migration of the a-Si interface, we note that triboepitaxy resembles grain-boundary migration in metals under shear[17–19] and that thermally induced migration of grain boundaries under anisotropic elastic strain is driven by the elastic energy density difference $\Delta E_{el}$ between grains.[20,21] We show in the following that $\Delta E_{el}$ also determines the growth direction in shear-induced migration. An arbitrary steady-state configuration from the MD trajectory in Figure 1f is selected and relaxed to $\tau = 0$ GPa and $T = 0$ K (Methods). This provides the starting structure for a quasistatic shear simulation where a finite shear stress $\tau_q > 0$ is imposed by a stepwise displacement of the upper rigid layer in the lateral direction while continuously relaxing the atomic positions. The dotted and solid curves in Figure 2b show the height dependence of the average potential energy per atom $E(z)$ at $\tau_q = 0$ and $\tau_q = \langle\tau\rangle$ from Figure 2a. A zoom-in (Figure 2c) reveals that the energy density of both crystals is degenerate at $\tau_q = 0$, while for $\tau_q > 0$ the upper crystal atoms have a higher energy ($\Delta E_{el} \approx 2.3$ meV). We propose that this gradient triggers the growth of the lower crystal.

We substantiate the existence of a clear correlation between $\Delta E_{el}$ and the growth direction by additional sliding simulations where we rotate the sliding direction by an angle $\alpha \in (0°, 90°]$ (Figure 2d, inset). Due to the symmetry of both crystals the system is inversed at $\alpha = 90°$, where the shear directions of the lower and upper crystal are $[1\bar{1}0]$ and $[00\bar{1}]$ (equivalent to $[001]$), respectively. The lower (upper) crystal grows for $\alpha < 45°$ ($\alpha > 45°$) with approximately constant growth rate $\xi$. For $\alpha = 45°$, the two crystals slide along equivalent directions and the time-average $\langle\xi\rangle$ is zero. Importantly, $\Delta E_{el}(\alpha)$ evaluated at the respective $\tau_q = \langle\tau\rangle$ closely correlates with $\langle\xi(\alpha)\rangle$ (Figure 2d).

The condition $\Delta E_{el} \neq 0$ results from an anisotropic elastic response of two misoriented crystals. The elastic energy density in a crystal under shear stress $\tau$ is given by $E_{el} = \frac{1}{2}\tau^2/G^*$, where $G^*$ is the effective shear modulus. Because of crystal anisotropy, rotation changes $G^*$ in the sliding direction. Since $\tau$ is determined by plasticity of the a-Si phase and hence independent of this rotation, the change in shear modulus carries over directly to $\Delta E_{el}$ between the two crystals. Consequently, $\Delta E_{el}$ can also be calculated by applying a simple shear deformation to the simulation cell of the individual crystals while relaxing the atomic positions and evaluating the potential energy differences per atom at $\langle\tau\rangle$ (Figure 2d and Methods). Alternatively, $\Delta E_{el}$ from linear elasticity also shows good agreement (Figure 2d) and requires only the elastic constants $C_{11}, C_{12}$ and $C_{44}$ of the cubic crystal.

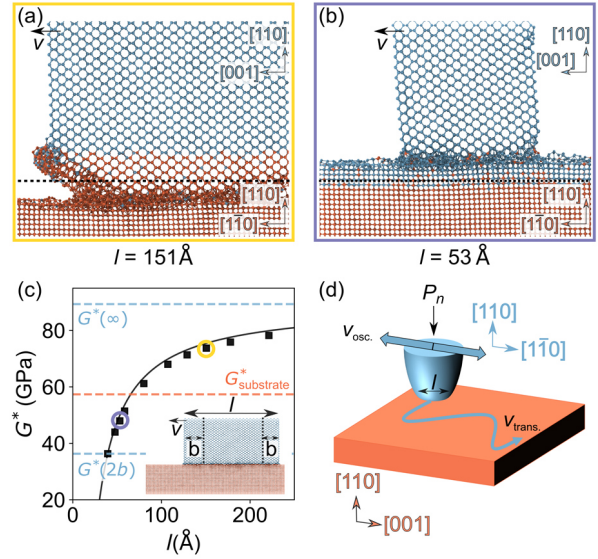

**Figure 3.** Triboepitaxy with a finite tip. (a, b) Snapshots of sliding simulations with tips sizes $l \approx 151$ Å and $l \approx 53$ Å at $s = 600$ nm along [001] of the tip and [1$\bar{1}$0] of the substrate. The horizontal dashed lines indicate the initial position of the interface between tip (blue) and substrate (red). (c) Effective shear modulus $G^*(l)$ of the tips ($l$ defined in the inset). Black squares: $G^*$ from quasistatic shear deformation. Dashed horizontal lines indicate $G^*$ of an infinite substrate ($G^*_{substrate}$) and tip ($G^*(\infty)$) and of a tip whose elastic response is completely determined by its surface ($G^*(2b)$). The surface thickness is estimated by $b \approx 19.68$ Å (Supporting Information S8). The solid curve shows the interpolation Eq. (1) between bulk and surface modulus. (d) Scheme for an experiment to deposit silicon nanostructures on (110) silicon substrates using (110) tips. High frequency oscillations with velocity $\vec{v}_{osc.}$ along [1$\bar{1}$0] of the tip and [001] of the substrate induce triboepitaxial growth. For writing, the tip is slowly translated with a velocity $\vec{v}_{trans.}(t)$ along arbitrary directions. Besides the oscillation direction $\hat{v}_{osc.}$ also surface orientation, $l$ and $P_n$ influence triboepitaxial growth (Supporting Information S6.4).

We now turn to triboepitaxial growth with finite-sized tips on flat substrates. Figures 3a,b show snapshots of Si(110) tips after sliding for $s \approx 600$ nm along their [001] direction on infinite Si(110) (along [1$\bar{1}$0]). According to Figure 2d, the tips should grow and, indeed, the tip with $l \approx 151$ Å follows this prediction (Figure 3a, Movie S3). Surprisingly, for $l \approx 53$ Å the substrate grows by ~4 atomic layers (Figure 3b, Movie S4). This can be ascribed to the strong size dependence of the effective elastic response of nanosized structures.[22] To determine the triboepitaxial growth direction, we have to consider the ordering of $G^*_{substrate}$ and $G^*_{tip}$. We perform quasistatic shear deformations of tips of various length $l$ to obtain their shear moduli $G^*(l)$ (Figure 3c, Methods). For large $l$, $G^*$ slowly converges to the value of an infinite (110) surface $G^*(\infty)$, while it rapidly decreases for $l < 20$ nm. In analogy to Ref. 23, we formulate an analytical expression for $G^*(l)$ by separating $G^*$ into a surface contribution $G^*(2b)$ defined by a surface thickness $b$ (Figure 3c, inset) and a bulk contribution $G^*(\infty)$. The analytical $G^*(l)$ given by the weighted average

$$G^*(l) = G^*(\infty) \cdot \frac{l-2b}{l} + G^*(2b) \cdot \frac{2b}{l} \qquad (1)$$

follows closely the quasistatic shearing results. Equation (1) suggests that an increase in the surface-to-volume ratio underlies the anomaly in triboepitaxial growth observed for the small tip in Figure 3b. Indeed, since $G^*(l = 151 \text{ Å}) > G^*_{substrate}$ the large tip grows, while $G^*(l = 53 \text{ Å}) < G^*_{substrate}$ determines epitaxial growth on the substrate. Thus, the tip size represents an additional control variable for triboepitaxy.

Finally, we propose an experimental setup for triboepitaxy (Figure 3d). A (110)-oriented nanoscale tip (blue) oscillates with velocity $\vec{v}_{osc.}$ along its $[1\bar{1}0]$ direction and rubs on a Si(110) substrate (red) along the $[001]$ direction. The choice of these orientations favor the growth of the substrate irrespective of the tip size. Further simulations reveal that a variety of surface orientations can be used and that the process is robust with respect to changes in $T$ and $P_n$ (Supporting Information S6, S7). Typically, in experiments $v_{osc.}$ is about 100 nm/s.[24,25] If $\langle \xi \rangle \sim 0.01$ (Figure 2d), we expect a growth velocity $\xi v_{osc.}$ of about 1 nm/s. The introduction of an additional translation $\vec{v}_{trans.}(t)$, with $v_{trans.} \ll v_{osc.}$, enables writing of arbitrary crystalline nanolines.

In conclusion, the triboepitaxy concept presented in this work lends itself to immediate experimental validation. Since triboepitaxial growth is simply governed by the elastic properties of the two sliding crystals, predictions on the crystal growth direction are straightforward and the process can be controlled by easily accessible parameters and implemented using existing technologies.[4] A first step in this direction could be made by using atomic force microscopy probes inside a transmission electron microscopy.[24] Adhesion experiments performed under these conditions suggest that surface-passivating species that prevent the formation of covalent bonds across the tribological interface are easily removed by sliding.[24] To further investigate the generality of the proposed mechanisms, simulations are underway to investigate triboepitaxy at sliding contacts between two differently oriented surfaces (Supporting Information S6.4). In such cases, due to the different elastic responses of the two crystals to normal pressure, the applied normal force can become an additional parameter to control crystal growth direction and rate. Since one of the main ideas behind triboepitaxy is that shear-induced plasticity can replace temperature to help atoms overcome energy barriers for recrystallization, the method could be tested on other materials showing shear-induced amorphization (*e.g.* diamond[9]) or materials on which solid-state epitaxy was applied successfully (*e.g.* germanium[11]). More generally, this study reveals a new way to apply tribological concepts in the context of nanolithography and nanofabrication. While mechanical scanning-probe techniques are usually based on nanomachining,[4] *i.e.* removal of material from a surface, we propose that tribologically induced phase transitions can be exploited for direct deposition of nanostructures or selective microstructural modification of a surface.

**Computational Methods**

**Interatomic potentials.** All atomistic simulations described in the article are performed with a screened version of the bond-order potential by Kumagai and coworkers.[12,13] Simulations employing different interatomic potentials that can describe shear-induced amorphization/crystallization processes and elastic anisotropy deliver analogous results (see Supporting Information S1 for a comparison to the Stillinger-Weber potential[14] and a more a recent potential based on machine learning of density-functional-theory results (Gaussian Approximation Potential, GAP)[15]).

**Molecular dynamics sliding simulations.** Figure 1 shows the setup for the sliding simulations of two periodic crystals. We do not consider passivation of the silicon surfaces, with *e.g.* H atoms, because experiments show that atoms that passivate the surface are easily removed during sliding and covalent bonds between silicon crystals form as a result.[24] After initial equilibration of normal pressure and temperature (300 K), a rigid layer of atoms consisting of the topmost 3.5 Å of the upper slab is driven at a constant velocity $v = 10$ m/s, while the 3.5-Å-thick bottom layer of the lower slab is kept fixed. The 5-Å-thick layers next to these two rigid layers are thermalized using a Langevin thermostat with a time constant of 0.1 ps. To minimize the effect of the thermostat dissipation on the shear stress, thermalization is performed only in the direction normal to the sliding and to the loading directions. For the simulations of Figure 2d, where the sliding direction is varied by an angle $\alpha$, the thermalization is instead performed in all three spatial directions. In this case, the velocity of the centre of mass of the thermalized slabs is subtracted from the atomic velocities in the calculation of the temperature. The normal pressure is applied using the barostat described in Ref. 26, which is specifically designed to provide realistic mechanical boundary conditions for periodic sliding systems. The equations of motion are integrated using a time step of 0.5 fs. Periodic-boundary conditions are applied along the directions perpendicular to the normal pressure. An overview of the system sizes is provided in the Supporting Information S2.

In general, in order to fit two crystals with different sliding direction in the same periodic simulation cell, the crystals have to be strained along the directions of the sliding plane to align their periodicities. This spurious strain changes the elastic response of the crystals and thus affects $\Delta E_{el}$. To mitigate this effect, we choose multiples of the unit cells such that the necessary strain for the alignment is minimal compatibly with an affordable computational cost. Moreover, we distribute the strain equally in both crystals. Additional tests shown in the Supporting Information S3 confirm that triboepitaxy is not affected by mismatch strain.

**Computation of the vertical positions $h_1$, $h_2$ of the amorphous-crystal interfaces.** For the computation of the position and thickness of the amorphous region, *i.e.* the determination of the two amorphous-crystal interfaces $h_1$ and $h_2$, we refer to our previous publication.[10] Each atom is categorized as crystalline or amorphous using OVITO's structure identification algorithm for diamond lattice.[27,28] We then divide the system in 1.5-Å-thick horizontal slices and determine the ratio between crystalline atoms and the total number of atoms in each slice. The ratios are averaged over 100 atomic configurations (corresponding to a sliding distance of 0.5 nm for $v = 10$ m/s). In this way, we obtain a ratio $\theta_{diam}$ as a function of $z$, where $z$ is the coordinate normal to the sliding plane. $h_1$ and $h_2$ are then obtained by fitting the function

$$\theta(z) = B + A[f_{FD}(z - h_1, \Delta z) + f_{FD}(h_2 - z, \Delta z)],$$

to $\theta_{diam}$. $f_{FD}$ is the Fermi-Dirac function $f_{FD}(z, \Delta z) = \frac{1}{1+e^{z/\Delta z}}$ and $\Delta z = 1$ Å. $A$, $B$, $h_1$ and $h_2$ are free parameters of the fit.

**Quasistatic shear simulations of silicon/silicon interfaces.** To determine $\Delta E_{el}$ for the periodic sliding simulations (Figure 2c and "Quasistatic shear" in Figure 2d), we select an arbitrary steady-state configuration along the trajectory (*i.e.* $\Delta h = \Delta h_{eq}$). We relax this configuration with the constraints that the lower rigid layer remains fixed while the upper rigid layer can move along the sliding direction in order to satisfy $\tau = 0$. Afterwards, we move the upper rigid layer in increments of 0.05 Å along the sliding direction. The whole system, excluding the rigid atoms, is relaxed after each displacement and the energy per atom as well as the shear stress on the system $\tau_q$ are measured. As explained in the Supporting Information S5, the choice of the particular steady-state atomic configuration does not significantly influence the value of $\Delta E_{el}$.

**Simple shear deformation of silicon single crystals.** Since in the periodic simulations $\Delta E_{el} \neq 0$ is the result of an anisotropic elastic response of the crystals, $\Delta E_{el}$ can also be calculated by applying a simple shear deformation to the two single crystals separately ("separate crystals" curve in Figure 2d). For each $\alpha$ we use two periodic bulk silicon crystals (consisting of 108 atoms each). One crystal has the orientation of the upper crystal and the other one has the orientation of the lower crystal. Both crystals are deformed by increasing the tilt of their simulation cell in the respective sliding direction in increments of 0.06 Å. The atomic positions are relaxed after each deformation. As soon as the shear stress $\tau_s$ on the simulation cell has reached the time-averaged shear stress $\langle \tau \rangle$ of the MD sliding simulation, we evaluate the energy difference per atom $\Delta E_{el}$ between the two separated crystals.

**Triboepitaxy simulations with finite tips.** To understand if and how finite-size effects can affect triboepitaxial growth, we perform simulations of a nano-sized Si tip rubbing on an infinite Si substrate. For simplicity, we choose a cuboid-shaped tip with dimensions $l \times 38 \text{ Å} \times 71 \text{ Å}$, where $l$ denotes the length of the tip in the sliding direction. We consider two tips with $l \approx 53$ Å and $l \approx 151$ Å. The size of the substrate is $127 \text{ Å} \times 38 \text{ Å} \times 46 \text{ Å}$ in the former case and $250 \text{ Å} \times 38 \text{ Å} \times 46 \text{ Å}$ in the latter case. Periodic boundary conditions are applied along the Cartesian directions lying in the interface plane. The topmost 3.5-Å-thick region of the tip and the lowest 3.5-Å-thick region of the substrate are kept rigid. Due to increased total system sizes (19253 and 44320 atoms for the two tip sizes) compared to the fully periodic systems used in the previous simulations and to the long simulation times required to observe triboepitaxy, we use a sliding velocity $v = 30$ m/s. This is applied to the top rigid layer after initial temperature and pressure equilibration (see Supporting Information S4 for a discussion of the influence of the sliding velocity). The system is equilibrated at $P_n = 0$ using the pressure-coupling algorithm described in Ref. 26. Atoms within 5-Å-thick regions next to the rigid layers are thermostated to dissipate the generated heat using a Langevin thermostat with target temperature 300 K and time constant 0.1 ps. In the calculation of the temperature, the velocity of the center of mass of each thermostated region is subtracted from the atomic velocities. The equations of motion are integrated using a time step of 0.5 fs.

**Calculation of the effective shear moduli $G^*$.** To calculate the effective shear moduli $G^*$ of the substrate and of the tip as a function of $l$ (Figure 3c), we perform quasistatic shear simulations of the tip and the substrate individually. For this purpose, we keep the outermost 3.5-Å-thick top and bottom atomic layers of each system fixed. After initial relaxation, the upper rigid layer is moved in increments of 0.05 Å along the shear direction. After each displacement, the remaining non-rigid atoms are relaxed and the shear stress is measured. The slope of the resulting stress-strain curve yields $G^*$. Here, strain is defined as the displacement of the upper rigid layer divided by the height of the system.

**Simulation Software.** Molecular dynamics and quasistatic shear simulations were carried out with LAMMPS code,[29] which was interfaced to the Atomistica library (github.com/Atomistica/atomistica) for the screened Kumagai potential and to the QUIP package (github.com/libAtoms/QUIP) for the GAP potential. For the simple shear deformations of silicon single crystals we used ASE[30] in combination with the ASE-Atomistica interface. Pre- and post-processing were carried out with ASE.[30] OVITO[27] was used for the visualization of atomic structures and for the structure identification of diamond lattices.[28]

**Author Contributions**

All authors conceived the research and wrote the manuscript. T.R. performed the simulations, data analysis and visualization. G.M., L. P. and M.M. supervised the research.

**Acknowledgments**

We are grateful to William Curtin for helpful discussions and to Andreas Klemenz and Hiroshi Uetsuka for initial contributions to the project. We gratefully acknowledge the Gauss Centre for Supercomputing e.V. (www.gauss-centre.eu) for funding this project by providing computing time through the John von Neumann Institute for Computing (NIC) on the GCS Supercomputer JUWELS,[31] at Jülich Supercomputing Centre (JSC). We also acknowledge computing time by the state of Baden-Württemberg through bwHPC and the DFG (grant no. INST 39/963-1 FUGG, bwForCluster NEMO). L.P. acknowledges support from the DFG (grant PA 2023/2).

**Supporting Information**.

Supporting Information: S1 Interatomic potentials; S2 Overview of the systems used for the simulations of periodic interfaces; S3 Role of strain due to lattice mismatch in periodic simulations; S4 Dependence of triboepitaxy on the sliding velocity; S5 $\Delta E_{el}$ as a function of the sliding distance; S6 Generalization to other surface orientations; S7 Robustness of triboepitaxy with respect to temperature and pressure variations; S8 Analytical model for the effective shear modulus $G^*(l)$ of the tip.

Movie S1: Interface migration of Figure 1e. The color of the atoms refers to the phase they belong to prior to sliding (blue for the upper, (110)-oriented crystal sliding along $[1\bar{1}0]$, red for the lower, (110)-oriented crystal sliding along [001], gray for amorphous atoms).

Movie S2: Trajectory of individual atoms (an extension of Figure 2a) that are initially part of the upper crystal and become part of the lower crystal upon triboepitaxial growth. The diagram tracks the $z$ position of the atoms (green, yellow, black). The shear stress $\tau$ is shown in violet. We note that abrupt motion of the atoms predominantly occurs during plastic slip phases.

Movie S3: Trajectory of individual atoms (Figure 3a) that are initially part of the flat substrate and become part of the tip ($l \approx 151$ Å) upon triboepitaxial growth. The included diagram tracks the $z$ position of the atoms (green, yellow, black), the shear stress $\tau$ is shown in violet. We note that abrupt motion of the atoms predominantly occurs during plastic slip phases.

Movie S4: Triboepitaxial growth on a $(110)[1\bar{1}0]$-oriented substrate using a tip with $l \approx 53$ Å sliding along the [001] direction (Figure 3b). The color of the atoms refers to the crystals the atoms belong to prior to sliding (blue for the tip, red for the substrate).

# Supporting Information

## S1 Interatomic potentials

All atomistic simulations described in the article are performed with a screened version of the bond-order potential by Kumagai and coworkers[1,2]. This interatomic potential can accurately describe elasticity, plastic processes and phase transitions of silicon[1–3]. At the same time its computational efficiency allows simulations at the time and length scales that are necessary to investigate triboepitaxial growth. The triboepitaxial process unveiled in the article relies on a correct description of shear-induced amorphization/crystallization processes and of elastic anisotropy. For this reason, it is important to mention that simulations employing other interatomic potentials that can suitably describe these processes deliver results that are analogous to those described in the article. Figure S1 summarizes the results of molecular dynamics (MD) simulations of triboepitaxial growth that we obtained using the Stillinger-Weber potential[4] and a more sophisticated machine-learning-based potential (Gaussian Approximation Potential, GAP)[5].

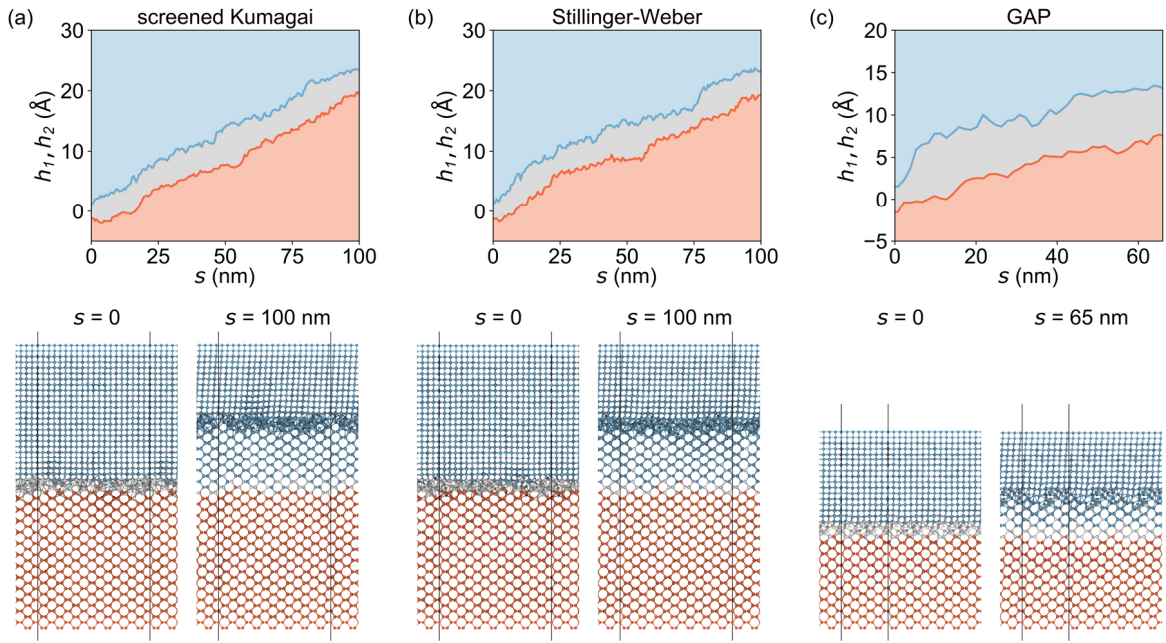

**Figure S1.** Triboepitaxy with different interatomic potentials. Position of the lower and upper amorphous-crystal interface $h_1$ and $h_2$ as a function of the sliding distance $s$ obtained with (a) the screened Kumagai potential[1,2], (b) the Stillinger-Weber potential[4], and (c) with the Gaussian Approximation Potential (GAP) "GAP_2017_6_17_60_4_3_56_165"[5]. The time-averaged growth

rates $\langle \xi \rangle$ are 0.025, 0.024 and 0.013, respectively. The system consists of two (110)-oriented crystals sliding along the [001] direction (lower crystal) and the [1$\bar{1}$0] direction (upper crystal) with either $v = 10$ m/s (a, b) or $v = 30$ m/s (c). The simulation temperature is $T = 300$ K and the normal pressure is $P_n = 0$. The system size for the Stillinger-Weber simulation is identical to the one used for the screened Kumagai simulation (see Table S1), but rescaled to the optimized Stillinger-Weber lattice constant 5.431 Å. For the GAP potential the system size is reduced to $15.9 \times 15.9 \times 67.3$ Å³ (850 atoms, 10 adatoms, lattice constant 5.461 Å) due to the higher computational cost. For the same reason, a higher sliding velocity is used in this case. The snapshots show the atomic configuration at different $s$, the colors of the atoms refer to the crystal the atoms belong to at $s = 0$.

## S2 Overview of the systems used for the simulations of periodic interfaces

| Crystal 1 | Crystal 2 | System size | | Strain crystal 1 | | Strain crystal 2 | | Atoms |
|---|---|---|---|---|---|---|---|---|
| | | $x$ (Å) | $y$ (Å) | $x$ (%) | $y$ (%) | $x$ (%) | $y$ (%) | |
| (001)[100] | (001)[100] | 38.00 | 38.00 | 0 | 0 | 0 | 0 | 7081 |
| (110)[1$\bar{1}$0] | (110)[1$\bar{1}$0] | 30.71 | 38.00 | 0 | 0 | 0 | 0 | 7165 |
| (110)[001]* | (110)[1$\bar{1}$0] | 38.20 | 27.01 | 0.5 | 0.5 | −0.5 | −0.5 | 5074 |
| (110)[001]* | (110)[1$\bar{1}$0] | 38.20 | 38.20 | 0.5 | −0.5 | −0.5 | 0.5 | 7165 |
| (001)[100] | (001)[110] | 38.20 | 38.20 | 0.5 | 0.5 | −0.5 | −0.5 | 7153 |
| (210)[001] | (210)[1$\bar{2}$0] | 48.71 | 48.71 | −0.3 | 0.3 | 0.3 | −0.3 | 11401 |
| (211)[21$\bar{5}$] | (211)[1$\bar{2}$0] | 46.58 | 46.58 | 4.4 | −4.1 | −4.1 | 4.4 | 11305 |
| (221)[11$\bar{4}$] | (221)[1$\bar{1}$0] | 34.55 | 34.55 | 0 | 0 | 0 | 0 | 5695 |
| (111)[1$\bar{1}$0] | (111)[11$\bar{2}$] | 46.31 | 26.73 | 0.5 | 0.5 | −0.5 | −0.5 | 5651 |
| (001)[100] | (221)[11$\bar{4}$] | 44.75 | 38.20 | 3.0 | 0.5 | −2.9 | −0.5 | 8257 |
| (001)[100] | (211)[1$\bar{2}$0] | 48.71 | 44.02 | −0.3 | 1.3 | 0.3 | −1.3 | 10849 |

*The results for the (110)[001] - (110)[1$\bar{1}$0] system are obtained using the setup with 5074 atoms, except for the those shown in Fig. 2d, where the symmetric setup ($x = y$) with 7165 atoms is used in order to ensure that the system is perfectly inversed at $\alpha = 90°$.

**Table S1.** Overview of the systems used for the simulations of periodic interfaces. The individual crystals are characterized using the notation $(h_s k_s l_s)[h_d k_d l_d]$, where the integer numbers between round brackets are the Miller indices of the sliding surface while those between square brackets indicate the sliding direction. $x$ refers to the length of the cell vector along the shear direction while $y$ refers to the direction in the interface plane normal to the shear direction. In each system, 25 adatoms are initially added to facilitate the phase transitions. Negative values indicate a compressive strain.

## S3 Role of strain due to lattice mismatch in periodic simulations

In general, in order to fit two crystals with different sliding direction in the same periodic simulation cell, the crystals have to be strained along the directions of the sliding plane to align their periodicities. This spurious strain changes the elastic response of the crystals and thus affects $\Delta E_{el}$. To mitigate this effect, we choose multiples of the unit cells such that the necessary strain for the alignment is minimal compatibly with an affordable computational cost. Moreover, we distribute the strain equally in both crystals (Table S1). Although the magnitude of the mismatch strain in our sliding systems is relatively low, we verify that the interface migration is not an artifact of the mismatch strain. We consider the $(110)[001]$ - $(110)[1\bar{1}0]$ system and perform two additional MD simulations: one in which the mismatch strain is fully applied to the upper crystal and another one for the inverse case. Figure S2 shows the position of the lower amorphous-crystal interface as a function of the sliding distance. Although imposing the spurious mismatch strain slightly increases the elastic energy of the strained crystal, this has no significant influence on triboepitaxial growth and in all three cases the growth direction and the growth rate are the same.

This test is especially interesting for the case in which the whole strain in accommodated in the $[001]$-oriented crystal. Here, without shear, the $[1\bar{1}0]$-oriented crystal is energetically more stable than the $[001]$-oriented crystal. If this interface migration was driven by temperature, without applied shear, we would expect a growth of the $[1\bar{1}0]$-oriented crystal. However, with increasing $\tau$, the sign of $\Delta E_{el}$ changes and the $[001]$-oriented becomes the most stable crystal and grows. To circumvent any artifact, the mismatch strain is equally distributed in both crystals.

An additional test shows that the sign of $\Delta E_{el}$ from quasistatic simulations of the tribosystem including strain (Fig. 2c) is identical to the sign of $\Delta E_{el}$ as obtained by simple shear deformations of the separate crystals (without any strain induced by lattice mismatch). This confirms that the energetic ordering of the crystals under shear is not affected by mismatch strain.

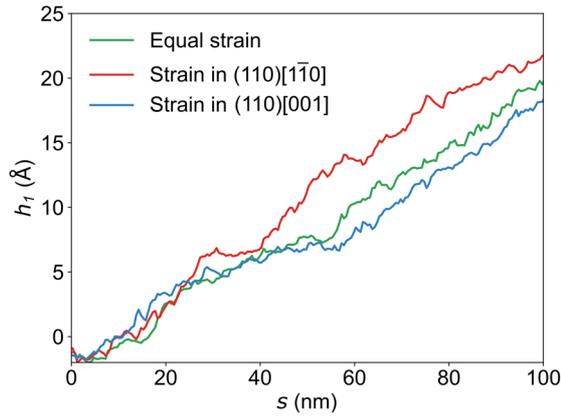

**Figure S2.** Role of strain due lattice mismatch in periodic simulations. Position $h_1$ of the lower amorphous-crystal interface as a function of $s$ for the $(110)[001]$ - $(110)[1\bar{1}0]$ sliding system for different distributions of the strain necessary to align the periodicities of the upper and the lower crystal. The strain is either equally distributed between both crystals ("Equal strain") or accommodated entirely in one of the crystals, while the other crystal remains in its optimal configuration ("Strain in $(110)[001]$ / $(110)[1\bar{1}0]$"). The normal load $P_n$ is 0, $T = 300$ K and $v = 10$ m/s.

## S4 Dependence of triboepitaxy on the sliding velocity

Irrespective of the sliding velocity, the magnitude of the crystal growth depends solely on the sliding distance $s$ for given temperature and normal pressure. Figure S3a shows the time evolution of $h_1$ for different sliding velocities. If $h_1$ is plotted as a function of the sliding distance $s = v \cdot t$ (Fig. S3b), all three curves approximately collapse into one. This was already observed for the mechanical amorphization of diamond[6] and silicon[3].

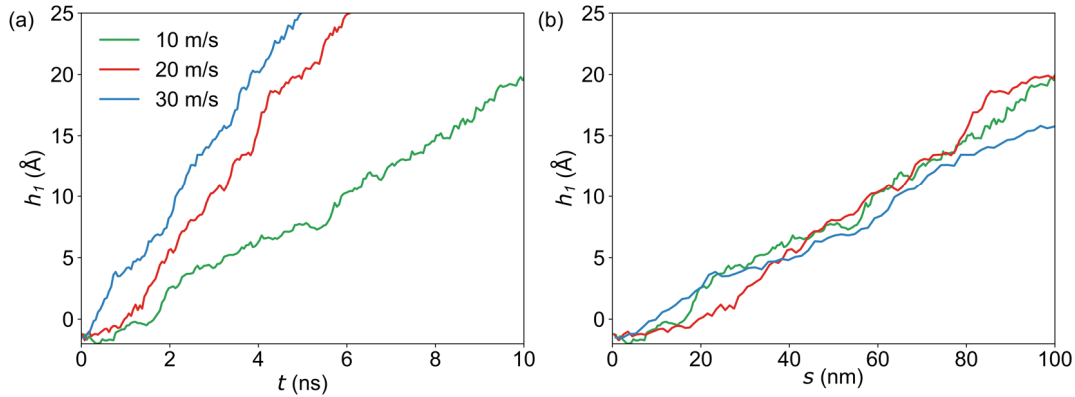

**Figure S3.** Dependence of triboepitaxy on the sliding velocity. Vertical position of the lower amorphous-crystal interface $h_1$ as a function of time $t$ (a) and of the sliding distance $s$ (b) for different sliding velocities $v$. The system consists of two (110)-oriented Si crystals. The lower crystal slides along [001], the upper crystal along [110]. The normal load $P_n$ is 0 and $T = 300$ K.

## S5 $\Delta E_{el}$ as a function of the sliding distance

In Fig. 2c and 2d, $\Delta E_{el}$ is evaluated by selecting a snapshot from the MD trajectory after the steady-state $\Delta h_{eq}$ is reached. In order to show that the choice of the particular steady-state snapshot does not affect the sign of $\Delta E_{el}$ (and even its magnitude), we perform the quasistatic shear simulations of Fig. 2c for 41 equidistant snapshots extracted from the MD trajectory between $s = 0$ and $s = 100$ nm. For each simulation, we determine three energy levels at $\tau_q = \langle \tau \rangle$ as shown in Fig. S4a: (i) the energy level of the lower crystal (dashed blue line); (ii) the energy level of the upper crystal (dashed red line); (iii) the maximal energy of the amorphous zone (dashed purple line). Figure S4b shows the evolution of all three levels as a function of $s$. The energy of the atoms in the amorphous zone undergoes rather large variations. However, no energy drift is observed as the position of the amorphous region changes. This is in agreement with the fact that the properties of the amorphous shear region remain statistically unaltered irrespective of its position. The energy levels of the crystals are very well defined and barely change in time. As a result, the sign of $\Delta E_{el}$ does not depend on the particular choice of the sliding distance during the steady-state phase in which $\Delta E_{el}$ is measured. As a matter of fact, even the magnitude of $\Delta E_{el}$ does not vary significantly during sliding.

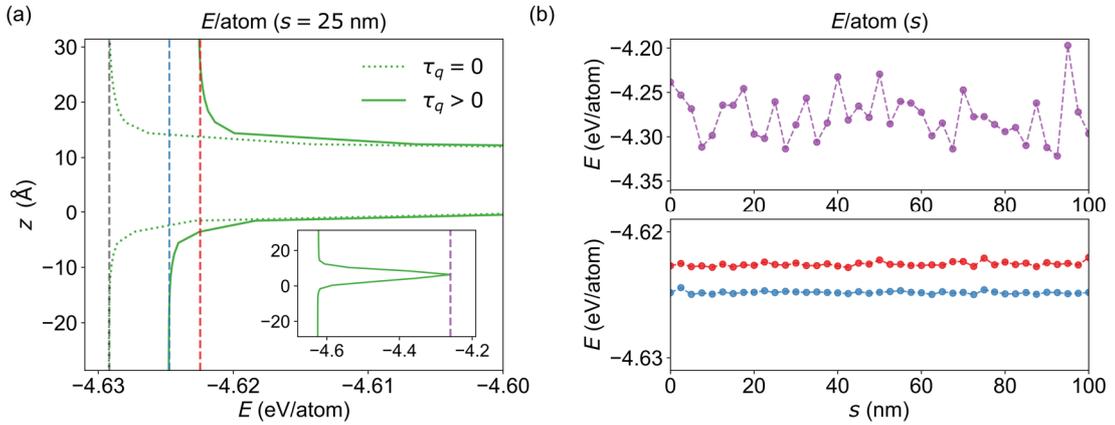

**Figure S4.** $\Delta E_{el}$ as a function of the sliding distance. (a) Averaged energy per atom $E$ for the (110)[001] - (110)[110] system as a function of the $z$ coordinate for $\tau_q = 0$ (dotted line) and for $\tau_q > 0$ (solid line) as obtained by quasistatic shear simulations. The initial configuration for this calculation is the atomic configuration at $s = 25$ nm (Fig. 1f). The gray dashed line is a guide to the eye highlighting that at $\tau_q = 0$ both crystals are energetically degenerate. The dashed blue line refers to the energy level of the lower crystal under shear, the dashed red line to the one of the upper crystal. The inset is a zoom-out, where additionally the maximal energy density of the amorphous zone is marked with a dashed purple line. Panel (b) shows the evolution of these energy levels as a function

of the sliding distance $s$. All energies are determined at the average shear stress $\langle\tau\rangle$ of the dynamic simulation.

## S6 Generalization to other surface orientations

### S6.1 Shear-induced amorphization and recrystallization for (001)-oriented crystals

To show that concurring amorphization and recrystallization processes also occur for other surface orientations and give rise to a steady-state with constant $\Delta h_{eq}$, we reproduce the results described in Fig. 1 for a sliding contact between two (001)-oriented surfaces (Fig. S5). Despite different amorphization/crystallization kinetics, the results are analogous to those we obtained for two (110)-oriented crystals. In particular:

- Upon sliding with different external loads, the symmetric system (i.e. same sliding direction for both crystals) undergoes mechanical amorphization and the thickness $\Delta h$ of the interfacial disordered layer increases with increasing $P_n$;

- $\Delta h$ grows with increasing sliding distance until it saturates at a constant value $\Delta h_{eq}$;

- $\Delta h_{eq}$ is a characteristic feature of the system under shear for given normal load and temperature;

- Continuously ongoing crystallization and amorphization events lead to a stochastic movement of the amorphous-crystal interfaces for the symmetric system;

- When the lower crystal is rotated such that the crystallographic sliding directions of the crystals are different, one of the two crystals (the lower in these examples) grows at the expenses of the other one.

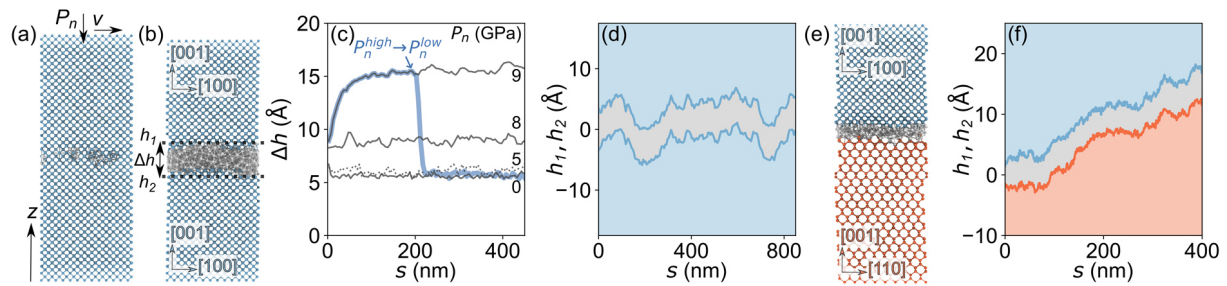

**Figure S5.** Shear-induced amorphization and recrystallization for (001)-oriented surfaces. Same as Fig. 1, but for (001)-oriented crystals. Both crystals slide along the [100] direction in (a)-(d), while in (e)-(f) the lower crystal slides along the [110] direction. (a) Initial crystals under normal load $P_n$

before sliding along the same [100] direction with $v = 10 \, ms^{-1}$. (b) After shearing at $P_n = 8$ GPa for a sliding distance $s = 300$ nm. Blue and gray spheres represent Si atoms in the crystalline and amorphous regions, respectively. (c) a-Si thickness $\Delta h(s, P_n)$ as a function of $s$ for different $P_n$. The dotted line refers to $P_n = 0$. The blue curve shows $\Delta h(s)$ when the initial $P_n^{high} = 9$ GPa is reduced to $P_n^{low} = 5$ GPa at $s = 200$ nm. (d) Vertical positions $h_1(s)$ and $h_2(s)$ of the two amorphous-crystal interfaces for $P_n = 5$ GPa. Gray and blue colors denote the amorphous and crystalline regions, respectively. (e) Shearing of differently oriented crystals: the upper crystal slides along the [100] and the lower crystal (in red) along the [110] direction. (f) $h_1(s)$ and $h_2(s)$ for the system in (e) with $P_n = 0$.

## S6.2 Triboepitaxy on other surface orientations

To assess the generality of the triboepitaxial growth process we repeat the periodic simulations of Fig. 1e,f for additional surface orientations. For (111)-oriented interfaces amorphization is strongly suppressed under the considered pressures due to the high stability of the shuffle plane[1] (similar to the diamond case[6]) and we do not observe interface migration.

For (001), (210), (211) and (221) surface orientations, tribo-interfaces consisting of two crystals with the same orientation but sliding along different crystallographic directions are constructed as shown in Fig. S6a. For all cases, we observe the formation of an interfacial amorphous layer with a constant height that migrates upwards with an approximately constant rate $\xi$ (Fig. S6b). Figure S6c shows the energy density difference $\Delta E_{el}$ between the upper and the lower crystal as a function of the shear stress $\tau_s$. We find that for any $\tau_s > 0$ the upper crystal is energetically less stable than the lower crystal and that the growth direction correlates with the sign of $\Delta E_{el}$. Apart for one case (red line), we also observe a correlation between $\Delta E_{el}$ and $\xi$. We note, however, that different amorphization/crystallization kinetics on different crystallographic orientations[7,8] are likely to affect the growth rate and, therefore, the correlation between growth rate and elastic energy difference.

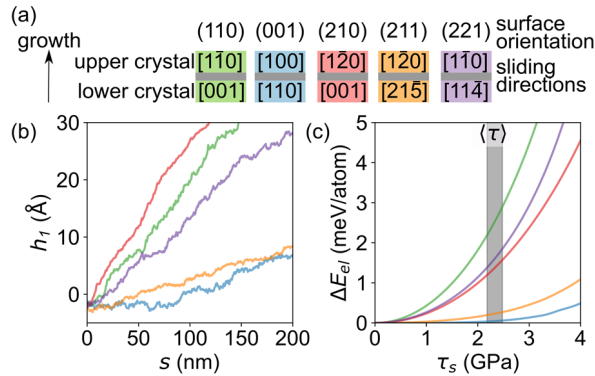

**Figure S6.** Elastically driven triboepitaxy for various surface orientations. (a) Two crystals with identical surface orientations slide along different crystallographic directions. (b) Vertical position $h_1(s)$ of the lower amorphous-crystal interface ($P_n = 0, T = 300$ K, $v = 10$ m/s). (c) Elastic energy difference $\Delta E_{el}$ as a function of the shear stress $\tau_s$ as determined by simple shear deformations of the separated silicon crystals. The shaded area indicates the range of the average shear stress $\langle \tau \rangle$ in the sliding MD simulations. The color code of (a) applies to (b) and (c).

### S6.3 Suitable orientations for triboepitaxial growth in an experiment

The most effective way to grow crystalline structures on a silicon substrate using the experiment suggested in Fig. 3d is to use tip/substrate systems in which both the finite size effects of the tip and the bulk elastic anisotropy favor the growth of the substrate crystal. One such example is the system presented in the article: both surfaces have a (110) orientation; the oscillatory motion for the tip is along the $[1\bar{1}0]$ direction and the equivalent $[\bar{1}10]$ direction, while the oscillatory motion for the substrate occurs along the $[001]$ and the equivalent $[00\bar{1}]$ direction.

It is important to point out that triboepitaxy is not restricted to these specific crystallographic orientations. Here, we discuss other suitable orientations for tip/substrate systems, where both the tip and the substrate have the same crystallographic surface orientation.

The case with perhaps the highest relevance for electronic applications is the triboepitaxial growth on (001) surfaces. One possibility to grow such surfaces is to use a (001)-oriented tip with oscillations along the $[100]$ and $[\bar{1}00]$ directions for the tip and along the $[110]$ and $[\bar{1}\bar{1}0]$ directions of the substrate. The elastic anisotropy alone leads to a rather slow growth rate of about $\langle \xi \rangle \approx 0.006$ if compared to other surface orientations (see Fig. S6b). Nonetheless, with the typical experimental velocities $v_{osc.} \approx$ 100 nm/s [9,10] significant growth velocities $\xi v_{osc.} \approx$ 0.6 nm/s can be achieved, which could be further increased by finite size effects. Alternatively, we find in preliminary simulations of silicon contacts with different surface orientations that the periodic $(001)[100]$ - $(110)[1\bar{1}0]$ system shows a higher rate $\langle \xi \rangle \approx 0.02$ and could also be used to grow (001) surfaces.

The so far discussed low-index surface orientations are symmetric with respect to an inversion of the sliding direction. This is not necessarily the case for higher index surfaces. Therefore, to be able to efficiently perform triboepitaxial growth on such surfaces using the experiment in Fig. 3d, one should make sure that the growth occurs in both sliding directions during oscillatory motion. Figure S7 shows how the energy per atom as a function of $\tau_s$ changes upon inversion of the deformation direction for the higher index surface orientations considered previously (Fig. S6).

For the (211) orientation (Fig. S7a) we find that an inversion of the oscillation direction can lead to an inversion of $\Delta E_{el}$. To avoid this, the suggested experiment should be performed by sliding a (211)-oriented tip along the $[1\bar{2}0]$ direction when the substrate slides along the $[21\bar{5}]$ direction. When the sliding direction is inversed, the tip slides along the $[\bar{1}20]$ direction and the substrate slides along the $[\overline{215}]$ direction. For the (210) and (221) orientations (Fig. S7b and c), the sliding inversion does not affect the sign of $\Delta E_{el}$. Triboepitaxy on (210)-oriented substrates is possible if the oscillation motion is along the tip's $[\bar{1}20]$ and $[1\bar{2}0]$ directions and the substrate's $[001]$ and $[00\bar{1}]$ directions. To grow structures triboepitaxially on (221)-surfaces, a (221)-oriented tip sliding along the $[1\bar{1}0]$ and the $[\bar{1}10]$ direction can be used, while the substrate slides along $[11\bar{4}]$ and $[\overline{11}4]$.

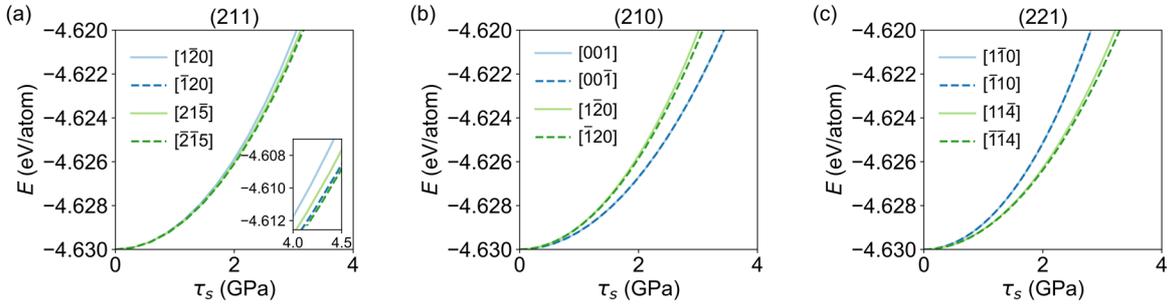

**Figure S7.** Effect of sliding direction inversion on $\Delta E_{el}$. Energy per atom $E$ as a function of the shear stress $\tau_s$ obtained for periodic silicon crystals deformed via simple shear. The title of each panel indicate the shear plane. The directions in the legends show the crystallographic shear directions. The inset in (a) is a zoom-in to reveal the ordering of the curves.

### S6.4 Triboepitaxy for interfaces between crystals with different surface orientation: Pressure as an additional parameter to control the growth direction

All simulations described in the article consider interfaces consisting of surfaces with identical crystallographic orientations. Preliminary simulations show that triboepitaxy can also be observed by pairing surfaces with different crystallographic orientations. In this context, the combination of crystals with very different elastic responses to normal load is particularly interesting, as this directly influences $\Delta E_{el}$.

Based on simple shear deformations of crystals under compressive strain, we identify two systems that show an inversion of $\Delta E_{el}$ upon an increase of $P_n$: A $(001)[100]$ - $(221)[11\bar{4}]$ sliding system and a $(001)[100]$ - $(211)[1\bar{2}0]$ sliding system. In both cases, for $P_n = 0$ the $(001)[100]$ crystal grows (dark green curves in Fig. S8a,b) – as one would expect from the sign of $\Delta E_{el}$ (Fig. S8c). With increasing normal load, $\Delta E_{el}$ and the migration rate $\langle \xi \rangle$ decrease until both quantities eventually become negative and the crystal growth direction is inverted (red curves in Fig. S8a,b). This reveals that due to different responses of the two crystals to the compressive load, the latter can be used to control the growth direction.

Figure S8c shows that $\Delta E_{el}$ and the migration rate $\langle \xi \rangle$ are correlated and that in the $(001)[100]$ - $(221)[11\bar{4}]$ system the growth direction is fully determined by the sign of $\Delta E_{el}$. However, for the $(001)[100]$ - $(211)[1\bar{2}0]$ case, the $\xi(\Delta E_{el})$ curve does not pass through the origin. While in the limit cases of high and no normal load the growth direction is consistent with the sign of $\Delta E_{el}$, there are intermediate values of the normal load, for which the signs of $\Delta E_{el}$ and $\langle \xi \rangle$ differ. The origin of this offset is not entirely clear. As shear stress and normal load have opposite effects on $\Delta E_{el}$, for a given value of the normal load, the sign of $\Delta E_{el}(\tau_q)$ can change with increasing $\tau_q$. In this case, the simplification to consider $\Delta E_{el}$ at the average shear stress $\langle \tau \rangle$ only is no longer meaningful. An energetic evaluation should probably rather be performed at every plastic deformation. Moreover, differences in amorphization/crystallization kinetics at different surfaces orientations[7,8] are likely to play an active role in the process.

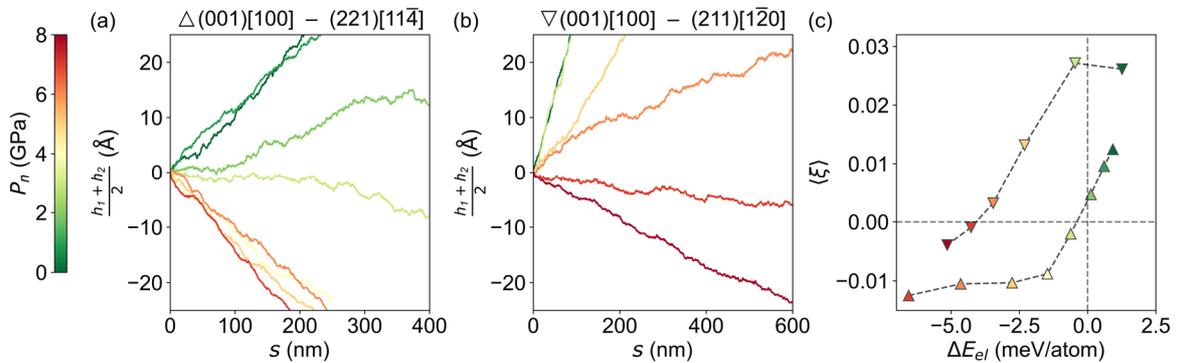

**Figure S8.** Inverting the growth direction with normal load for interfaces between crystals with different surface orientation. Position of the interfacial amorphous region $(h_1 + h_2)/2$ for a (001)[100] crystal sliding against a (221)[11$\bar{4}$] crystal (a) and for a (001)[100] crystal sliding against a (211) [1$\bar{2}$0] crystal (b) as a function of the sliding distance $s$. The temperature is set to 300 K, $v = 10$ m/s and the normal pressure is varied as indicated by the legend on the left. A positive $(h_1 + h_2)/2$ corresponds to the growth of the (001)[100] crystal. (c) Migration rate $\langle \xi \rangle$, determined by the slopes of the curves in (a) and (b), as a function of $\Delta E_{el}$, as determined by quasistatic shear simulation at the average shear stress $\langle \tau \rangle$ of the respective simulation.

## S7 Robustness of triboepitaxy with respect to temperature and pressure variations

To understand how triboepitaxy depends on temperature and normal load, we perform two additional series of simulations using the tribological systems shown in Fig. S9a.

Firstly, we reduce the target temperature of the thermostats from 300 K (Fig. S9b) to 0 K (Fig. S9c) while maintaining the same normal load $P_n = 0$. In the latter case, the temperature at the interface is typically around 30 - 40 K. Since the triboepitaxial process is mechanically driven, it should take place even at very low temperature as long as the shear rate in the interface amorphous region is finite (the average shear rate is given by $v/\Delta h$). For all interfaces, we observe that the decrease in temperature does not affect the growth direction, while it causes a decrease in the growth rate.

Secondly, we increase the normal load from $P_n = 0$ to $P_n = 5$ GPa (Fig. S9d), while keeping the temperature constant at 300 K. We find that the growth rates are not significantly affected by the pressure variation. Since both crystals in each interface have the same surface orientation, the elastic response to normal load is approximately identical. This can change drastically when crystals with different surface orientations are paired (a discussion can be found in Section S6.4).

The important aspect is that in all examined cases the growth direction is determined by the sign of $\Delta E_{el}$ as calculated by quasistatic shear simulations in spite of substantial variations of temperature and normal pressure. In conclusion, the reported phenomena are not restricted to a narrow window of specific operation conditions and we expect them to be robust with respect to temperature and normal load variations.

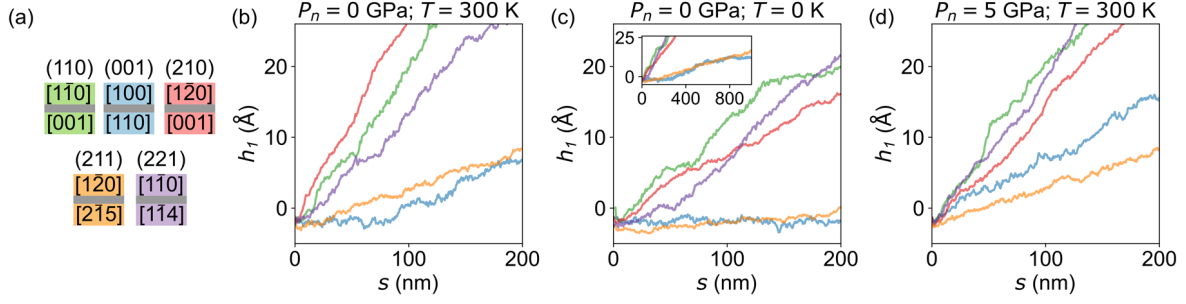

**Figure S9.** Temperature and pressure dependence of triboepitaxy. Vertical position $h_1$ of the lower amorphous-crystal interface for the systems depicted in (a) as a function of the sliding distance $s$. Panel (b) is a reproduction of Fig. S6b. Panel (c) shows the results when the target temperature of the thermostat is reduced to $T = 0$ K. The inset depicts the same curves for a longer sliding distance to show that triboepitaxy occurs for all cases after sufficiently long sliding distances. In panel (d) the normal load is increased to $P_n = 5$ GPa with $T = 300$ K. The sliding velocity is $v = 10$ m/s for all simulations. The color code of (a) applies to (b)-(d).

## S8 Analytical model for the effective shear modulus $G^*(l)$ of the tip

The analytical expression for the effective shear modulus of the tip $G^*(l) = G^*(\infty) \cdot \frac{l-2b}{l} + G^*(2b) \cdot \frac{2b}{l}$ as a function of its length $l$ consists a of bulk contribution $G^*(\infty)$ and a surface contribution $G^*(2b)$. The surface thickness $b$ (Fig. 3c) is the thickness of the tip's surface region in which the elastic response differs from the bulk elastic response, i.e. the response of the infinite tip ($l = \infty$). In Ref. 11, where an analogous model is proposed for quartz plates, $b$ is set to 10 Å to separate the surface and bulk contributions to the effective Young's modulus of the system. To estimate an upper bound for $b$, we refer to the typical decay length of the stress/strain field ahead of an atomically sharp crack in a silicon crystal of about 20 Å [12]. Applying the constraint 10 Å < $b$ < 20 Å, and considering that $l$ can only be varied by multiples of the distance between equivalent surface planes (2.71 Å) to allow for the calculation of $G^*(2b)$, the best fit to the $G^*(l)$ values calculated by means of quasistatic shear deformation is provided by $b = 19.68$ Å (Fig. S10). Due to the simplicity of the analytic model, which just serves as a qualitative model for finite-size elastic effects, the resulting $b$ value is a qualitative estimate, albeit physically reasonable. This, however, does not affect the conclusions drawn in the article, which are uniquely based on the simulations results.

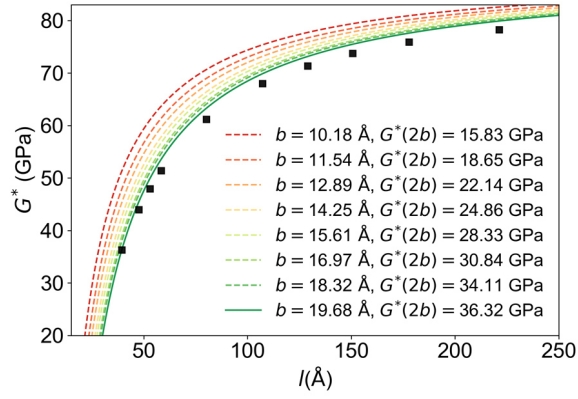

**Figure S10.** Effective shear modulus $G^*(l)$ of the $(110)[001]$ tip (Fig. 3c) as a function of its length $l$. Black squares show $G^*$ from quasistatic shear deformation. The curves show the interpolation $G^*(l) = G^*(\infty) \cdot \frac{l-2b}{l} + G^*(2b) \cdot \frac{2b}{l}$ between the shear modulus of the bulk region $G^*(\infty) = 89.34$ GPa and the shear modulus of the surface region $G^*(2b)$ for different values of the surface thickness $b$. The solid curve shows the best fit to the data points for $b \in [10, 20]$ Å.